\documentclass[epj]{svjour}

\usepackage{graphicx}
\usepackage{amssymb}
\usepackage{amsmath}
\usepackage{fancyhdr}
\def\makeheadbox{{%  remove EPJC header
 }}
\def\mytitle{My title} 
\def\myauthors{My name}  
\def\mytype{My type of session}
\def\mysession{My session}
\def\to{\rightarrow}

\def\bi{\begin{itemize}}
\def\ei{\end{itemize}}
\def\be{\begin{equation}}
\def\ee{\end{equation}}
\def\bea{\begin{eqnarray}}
\def\eea{\end{eqnarray}}

\def\lsim{\mathrel{\rlap{\lower4pt\hbox{\hskip1pt$\sim$}}
    \raise1pt\hbox{$<$}}}                % less than or approx. symbol
\def\gsim{\mathrel{\rlap{\lower4pt\hbox{\hskip1pt$\sim$}}
    \raise1pt\hbox{$>$}}}                % greater than or approx. symbol

\def\mytitle{Light MSSM Higgs Boson Scenario and its Test at Hadron Colliders} %Put your title here!
\def\myauthors{Alexander Belyaev}    %Put your name here!
\def\mytype{Contributed Talk}    
\def\mysession{Colliders - Higgs Phenomenology}
\newcommand{\ca}{c_\alpha}
\newcommand{\sa}{s_\alpha}

\begin{document}
\title{Phenomenology of Light MSSM Higgs Boson Scenario}
\author{Alexander Belyaev\inst{1}
\thanks{\emph{Email:} a.belyaev@soton.ac.uk}%
\thanks{the study has been done in 
collaboration with Qing-Hong Cao, Daisuke Nomura, Kazuhiro Tobe and  C.-P. Yuan~\cite{Belyaev:2006rf}}% \and
% Second author\inst{2}% etc
}                     % Do not remove
\institute{School of Physics and Astronomy, University of Southampton, Southampton, SO17 1BJ, U.K.}
%\institute{Physics Department, University of California Riverside, Riverside, CA 92521, USA}
%\institute{Theory Group, KEK 1-1 Oho, Tsukuba 305-0801, Japan}

\date{}
\abstract{
We have found that in the MSSM, the
possibility for the lightest CP-even Higgs boson  to be lighter
than $Z$ boson (as low as about 60 GeV) is, contrary to the usual
belief, not yet excluded by LEP2 Higgs search nor any direct
searches for supersymmetric particles at high energy colliders.
The Light Higgs boson scenario (LHS) is realised when
the $ZZh$ coupling and the decay branching ratio ${\rm Br}(h/A\to
b\bar{b})$ are simultaneously suppressed as a result of generic
supersymmetric loop corrections. Consequently, the $W^\pm H^\mp h$
coupling has to be large due to the sum rule of Higgs couplings to
weak gauge bosons and as we demonstrate,  the associate neutral
and  charged Higgs boson production process, $pp\to H^\pm h (A)$,
at the LHC
can completely probe the LHS.
 % end of PACS codes
} %end of abstract
\maketitle
%DO NOT REMOVE THIS LINE
%

%\section{Introduction}

While the Standard Model (SM) of particle physics is consistent
with existing data, there is a strong belief in a more complete
description of the underlying physics. Supersymmetry (SUSY), as a
good candidate for theory beyond the SM, solves principal
theoretical problems of the SM such as hierarchy and fine tuning,
as well as provides good dark matter candidate and potentially
solves the problem of baryongenesis.
In the minimal
supersymmetric standard model (MSSM)
the Higgs sector consists of {\it two} doublet fields $h_d$ and $h_u$
to generate masses for down- and up-type fermions, respectively,
and to provide an anomaly-free theory.
After spontaneous symmetry breaking, there remain
five physical Higgs bosons: a pair of charged
Higgs bosons $H^{\pm}$, two neutral CP-even scalars
$H$ (heavier) and $h$ (lighter), and a neutral
CP-odd pseudoscalar $A$.
Higgs potential is constrained
by supersymmetry such that all the tree-level Higgs boson
masses and self-couplings are determined by only two independent unknown
parameters, commonly chosen to be the mass of the CP-odd
pseudoscalar ($M_A$) and the ratio of  vacuum expectation
values  of neutral Higgs fields,
denoted as $\tan \beta\equiv \langle h_u\rangle /\langle h_d\rangle$.

The MSSM predicts a light neutral Higgs boson which is
lighter than $Z$-boson at the tree level.
However, large top quark and squark~(stop) loop contributions induce  significant
radiative correction to the Higgs quartic coupling, such that the lighter neutral Higgs
boson mass can be as large as
$130$ GeV~\cite{Okada:1990vk,Haber:1990aw,Ellis:1990nz,Barbieri:1990ja}.
The negative result of Higgs boson search at LEP2
via $e^+e^-\to Z h$ production channel imposes a lower
bound on the SM Higgs boson mass $M_h >114~{\rm GeV}$~\cite{Barate:2003sz},
and excludes significant
portion of MSSM parameter space.

The LEP2 collaborations
have performed analyses for the MSSM~\cite{unknown:2006cr}
using several benchmark scenarios
that were considered as typical cases for the MSSM parameter space.
The two complementary processes for MSSM Higgs boson search are $e^+e^-\to
Zh/Ah$~\cite{unknown:2006cr}, in which
 the first one occurs via $ZZh$ coupling $g_{ZZh}
=\sin(\beta-\alpha)(\equiv s_{\beta\alpha})$ while the second one  via
$ZAh$ coupling $g_{ZAh} = \cos(\beta-\alpha)$. The
obvious sum rule ($g_{ZZh}^2+g_{ZAh}^2=1$) puts strong constraints
on the mass and couplings of the MSSM Higgs boson $h$.
For all studied  benchmark scenarios at LEP2, $M_h$ below about
$90$ GeV is excluded~\cite{unknown:2006cr}.

In
this study, we propose a different region of the MSSM parameter space
which has not been previously studied  with deserved attention.
We call this possibility
light Higgs boson scenario (LHS),
in which the Higgs boson $h$ is lighter than the $Z$-boson
and the $ZZh$ coupling is small enough to be consistent with
the LEP2 data.

To satisfy the LEP2 constraint derived from the
production channel $e^+e^- \rightarrow Zh$ with $M_h<M_Z$, the
coupling $g_{ZZh}$ ({\it i.e.} $s_{\beta\alpha}$) has to be small.
Let us denote ${\cal M}^2$ as the $2\times 2$ squared-mass matrix of
the CP-even neutral Higgs bosons in the gauge eigenbasis $({\rm
Re}~h_d^0,{\rm Re}~h_u^0)$.
The mass eigenstates $(h, H)$  are given by the diagonalization of
the matrix ${\cal M}^2$ with the definition:
\begin{equation}
\label{hmix}
\left(
\begin{matrix}
   h \\
   H
\end{matrix}
\right)
=
 \left(
  \begin{matrix}
   -\sa & \ca \\
    \ca &  \sa
 \end{matrix}
\right)
\left(
\begin{matrix}
 {\rm Re}~ h_d^0\\
 {\rm Re}~ h_u^0
\end{matrix}
\right),
\end{equation}
where  $c_\alpha\equiv\cos\alpha~{\rm and}~s_\alpha
\equiv\sin\alpha$ (with $-\pi/2\leq\alpha\leq\pi/2$).

Denote $x\equiv {\cal M}^2_{11}-{\cal M}^2_{22}$ and $y\equiv{\cal
M}^2_{12}$, in terms of the components of
matrix ${\cal M}^2_{ij}$.
For relatively large $\tan\beta$ (as preferred by the LHS) and
$y/x\simeq 0$, we find $s_{\beta\alpha}\simeq
\frac{(|x|+x)^{1/2}}{\sqrt{2|x|}}$ which vanishes for $x<0$.
Therefore, conditions $y/x\simeq 0$ and $x<0$ provide small values
of $s_{\beta\alpha}$. The light Higgs boson $h$
mainly consists of $h_d^0$, and the neutral Higgs boson masses are
approximately given by $M^2_h\simeq {\cal M}^2_{11}$ and
$M^2_H\simeq {\cal M}^2_{22}$. This feature is different from the
usual scenarios in which $M^2_h\simeq {\cal M}^2_{22}$ and
$M^2_H\simeq {\cal M}^2_{11}$. As it is well-known, ${\cal
M}^2_{22}$ (i.e.~$h_u$-component) receives large positive
logarithmic correction from top and stop contributions.
This correction, which helps to significantly increase the mass of
$h$ in the usual scenarios, increases the mass of $H$ in the LHS
and changes the sign of $x$ value from positive (at tree level)
to negative when $M_A\sim M_Z$.
The condition $y/x\sim 0$ (needed for the LHS)
can only be satisfied in some regions of
MSSM parameter space which is studied below.
\begin{table}
\vspace*{-0.4cm}
\begin{tabular}{|l|l|l|l|l|l|l|}
%\begin{tabular}{|c|c|c|c|c|c|c|c|}
\hline
%point #92
 $\tan\beta$  & $M_{H^+}$  & $\mu$~~ &  $A_3$~   &  $M_1=M_2/2$~ & $M_3$     & $M_{Q}$\\
 $35$         & $135$      & $890$  &   $750$    &  $100      $  & $600$     & $330$  \\
\hline
\hline
\multicolumn{7}{|c|}
{$M_h=71$, $M_A=113$, $M_H=119$} \\
\multicolumn{7}{|c|}
{${\rm Br}(h/A/H\to b\bar{b})      = 0.65/0.64/0.03$} \\
\multicolumn{7}{|c|}
{${\rm Br}(h/A/H\to \tau\bar{\tau})= 0.25/0.34/0.54$} \\
\multicolumn{7}{|c|}
{$g_{ZZh}^2=0.006,~ g_{ZZH}^2=g_{H^+ W^- h}^2=0.994$} \\
\multicolumn{7}{|c|}
%cpy%
{$M_{\tilde\chi^0_1}=100$,
 $M_{\tilde\chi^+_1}=198$, $M_{\tilde{t}_1}=126$, $M_{\tilde{b}_1}=273$}\\
\multicolumn{7}{|c|}
{$\Delta\rho=6.7\times 10^{-4} $
%,\ \ $Br(b\to s\gamma)=3.57\times 10^{-4}$
} \\
\hline
\end{tabular}
\caption{\label{tab:bench}
The MSSM parameters (at the weak scale) of an LHS sample point. The
dimension of mass parameters is in unit of GeV.
$M_i(i=1,\ldots, 3)$, $M_{Q}$ and $A_3$ are gaugino masses, the universal soft-breaking sfermion mass
and universal trilinear A-term for the third-generation at the weak scale, respectively.
$M_{\tilde\chi^+_1}$, $M_{\tilde{t}_1}$ and $M_{\tilde{b}_1}$ are pole masses
for the lightest chargino, stop and sbottom, respectively.
}
\vspace*{-0.5cm}
\end{table}
As an illustration, we present in
Table~\ref{tab:bench} one LHS sample point
where the gaugino masses (with $M_2=2 M_1$),
the supersymmetric Higgs mass $\mu$-parameter ($\mu$),
the universal soft-breaking sfermion mass ($M_Q$),
and the trilinear A-term ($A_3$) for the third-generation at the weak
scale are all at (or below) TeV scale.
For our numerical analysis, we use CPsuperH
program~\cite{Lee:2003nt} and assume CP is conserved.
For the LHS sample point specified in Table~I,
%cpy%
$x>0$, $y/x\simeq -0.2$ and
$s_{\beta\alpha}\simeq 0.98$ at tree level.
After including radiative corrections,
 the Higgs mass matrix elements in
the effective potential become ${\cal M}^2_{11}\simeq (71.0~{\rm
GeV})^2,~{\cal M}^2_{22}\simeq (119.7~{\rm GeV})^2,$ and ${\cal
M}_{12}^2\simeq -(19.5~{\rm GeV})^2$, hence, $x<0$ and
$y/x\simeq 0.041$. (The mass of top quark is taken to be 172.5 GeV.)
 Consequently, we
obtain a small $s_{\beta \alpha}$ ($\simeq 0.069$). Note that in
the LHS, the lighter Higgs boson mass is close to its tree-level
value $M_h\simeq \sqrt{{\cal M}^2_{11}}\sim M_Z$ when $M_A\sim
M_Z$. This feature is qualitatively very different from
the commonly discussed MSSM scenarios in which $M_h$ receives large
radiative corrections. Moreover, the mass
of the heavier CP-even Higgs boson $H$ must receive large radiative
corrections to exceed about 114 GeV in order to agree
with  LEP2 data, since the  $ZZH$ coupling is close to the SM
value.

To find the allowed parameter space for the LHS with $\mu > 0$,
we scan the following set of MSSM parameters: {\small $\tan\beta~[1.1, 50]$,
%cpy%
$(M_{H^+}/{\rm TeV})$ $[0.1, 0.2]$, $(A_3/{\rm TeV})$ $[-2, 2]$,
$(M_{1}/{\rm TeV})$ $[0.05, 1]$, $(M_3/{\rm TeV})$ $ [0.05, 1]$,
$(M_{Q}/{\rm TeV})$ $[0.05, 1]$ and
$(\mu/{\rm TeV})$ $[0, 3M_Q]$}, within the range denoted in brackets.
Since a too large $\mu$-parameter induces not only the color breaking
vacuum in the general direction
of the scalar potential but also the fine-tuning in the Higgs mass parameter,
we require  $\mu$ to be less than $3M_Q$ in our analysis~\cite{Frere:1983ag,Claudson:1983et}.
Then, we check the LHS parameter space against the full set of the experimental
and theoretical constraints.
 They are:
(1) LEP2 $Zh/ZH$ and $Ah/AH$ constraints, cf. Tables 14 and 17 of
Ref.~\cite{unknown:2006cr}; (2) Chargino ($M_{\tilde{\chi}^+_1}$),
stop ($M_{\tilde{t}_1}$), sbottom ($M_{\tilde{b}_1}$) and gluino
($M_3$)
  mass limits:
%chargino
$M_{\tilde{\chi}^+_1}>103$~GeV~\cite{Yao:2006px},
%stop
$M_{\tilde{t}_1}>96$~GeV~\cite{Yao:2006px},
%sbottom
$M_{\tilde{b}_1} > 220$ GeV for $M_{\tilde{\chi}^0_1}<90$ GeV
and $M_{\tilde{b}_1} - M_{\tilde{\chi}^0_1} > 6$~GeV (where $M_{\tilde{\chi}^0_1}$ is the
lightest neutralino mass)~\cite{Abazov:2006fe}
or $M_{\tilde{b}_1} >100$ GeV for all other regions~\cite{Yao:2006px},
and
%gluino
$M_3 > 270$~GeV for $M_{\tilde{b}_1} < 220$~GeV
and $M_3 - M_{\tilde{b}_1} > 6$~GeV~\cite{Abulencia:2005us} or
$M_3>240$~ GeV for all other regions~\cite{Abazov:2006bj};
(3) electroweak constraint: one-loop stop
contributions to $\rho$-parameter $|\Delta\rho_{\rm stop}| < 2\times
10^{-3}$~\cite{Drees:1990dx};
(4) color breaking constraint:
$A_3^2<3(2M_{Q}^2+M_{h_u}^2+\mu^2)$ where $M_{h_u}$ is the soft-breaking mass
for Higgs $h_u$~\cite{Frere:1983ag,Claudson:1983et}.

\begin{figure}[htbp]
\vspace*{-0.3cm}
\includegraphics[width=0.5\textwidth]{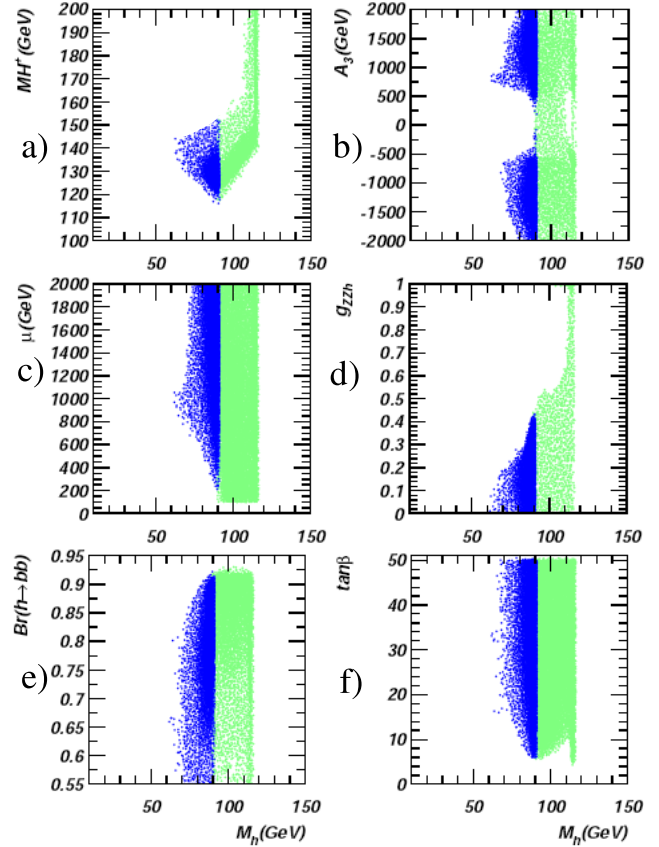}
\vspace*{-0.4cm}
\caption{\label{2d} Projected planes of scanned parameter space
indicating the LHS region in accord with experimental data. (See
detail explanation in the text.)}
\vspace*{-0.4cm}
\end{figure}

Our result is  shown in Fig.~\ref{2d} where blue (darker) and
green (lighter) color indicates allowed parameter space with $M_h<M_Z$
and $M_h>M_Z$, respectively.

Fig.~\ref{2d}a ($M_{H^+}$-$M_{h}$ plane) shows that LHS scenario
is realized for low values of charged Higgs boson mass:
$120~\mbox{GeV}<M_{H^+}<150$~GeV, indicating the non-decoupling
regime. Much lighter charged Higgses are excluded mainly by the
LEP2 Higgs search via the $Ah$ production channel. The scenario
requires intermediate-to-large values of the A-term and
$\mu$-parameter, $|A_3|>400$~GeV and $\mu\gsim 300$GeV (cf.
Figs.~\ref{2d}b and c) to make $g_{ZZh}$ small, as indicated in
Fig.~\ref{2d}d. On the other hand, a larger {\it positive} value
of the product \
$M_3\mu\tan\beta$ \ gives rise to larger {\it negative}
correction to the bottom Yukawa coupling $y_{hbb}$. This large
negative correction to $y_{hbb}$ is {\it non-universal} with
respect to the $\tau$ Yukawa coupling $y_{h\tau\tau}$ and leads to
a suppression in ${\rm Br}(h/A \to b\bar{b})$~\cite{Carena:1998gk}
large enough to avoid LEP2 constraint from the $Ah$ channel with
low $M_h$ values.
In the LHS parameter space,
${\rm Br}(h/A \to b\bar{b})$ can be suppressed down to about 50\%
(cf. Fig.~\ref{2d}e), and consequently ${\rm Br}(h/A \to
\tau\bar{\tau})$ is enhanced up to about 50\%, so that  $Ah$
channel is not observed: $b{\bar b}b{\bar b}$ decay mode is
largely suppressed, while $b{\bar b}\tau{\bar \tau}$ or $\tau{\bar
\tau}\tau{\bar \tau}$ signatures are not enhanced enough to
exclude 60 GeV$\lesssim M_h < M_Z$. Fig.~\ref{2d}e presents the
${\rm Br}(h\to b\bar{b})$--$M_h$ correlations. It is interesting
to note that the relatively large $\mu$-parameter simultaneously
suppress both $s_{\beta\alpha}$ and ${\rm Br}(h/A \to b\bar{b})$
to be consistent with the LEP2 data. We also note that
a lighter Higgs boson is preferred for a larger $\tan\beta$ value,
cf. Fig.~\ref{2d}f. It is worth mentioning that although the heavier
Higgs boson ($H$) couplings to vector bosons are SM-like, its couplings to
down-type fermions are further suppressed as compared to those of
$h$ and $A$ (cf. Table~\ref{tab:bench}).
Moreover, $M_A$ ranges from 90 to 120 GeV, which can be well approximated by
$M_A=\sqrt{M^2_{H^+}-M_W^2}$.

Since in the LHS, $g_{ZZh} (=s_{\beta\alpha})$ is suppressed,
 {\small $H^+W^- h$} coupling is inevitably enhanced due to the sum rules in
Higgs boson couplings to weak gauge bosons, i.e.,
$g_{ZZh}^2+g_{H^+ W^- h}^2=1=g_{H^+ W^- A}^2$. In this case the 
$q\bar{q}'\to H^\pm h (A)$ production via $W$ boson exchange could
be sizable with the production cross section $\sim$ 10 fb at the
Tevatron and $\sim$ 100 fb at the LHC for $M_{h/A}\sim 100$ GeV
~\cite{Kanemura:2001hz,Cao:2003tr}.
\begin{figure}[htb]
\vspace*{-0.3cm}
\includegraphics[width=0.5\textwidth]{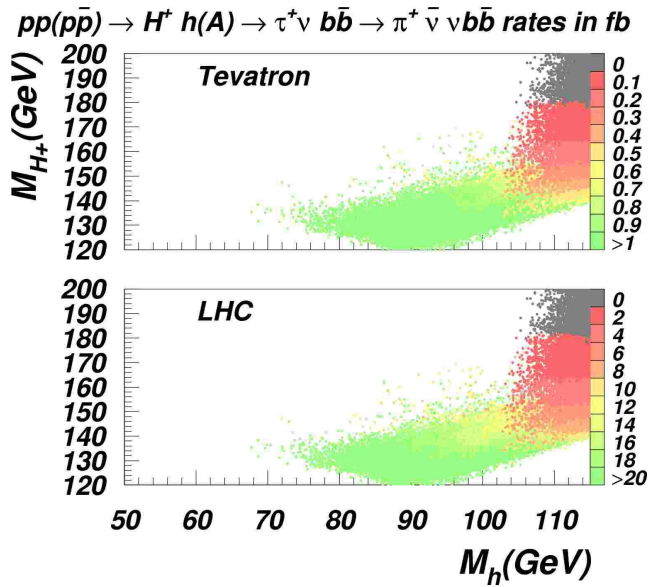}
\vspace*{-0.4cm}
\caption{\label{fig:cs} Rates for $p{\bar p},\, pp
\to H^+ h(A)\to\tau^+\nu b\bar{b}\to \pi^+\bar{\nu}\nu b\bar{b}$
signature at the Tevatron and the LHC.}
\vspace*{-0.2cm}
\end{figure}
In Fig.~\ref{fig:cs} we present the inclusive cross section of the
$p{\bar p},\, pp \to H^+ h(A)\to\tau^+\nu b\bar{b}\to
\pi^+\bar{\nu}\nu b\bar{b}$ signature at the Tevatron and the
LHC in the $M_{H^\pm}$-$M_h$ plane. For simplicity, we have
combined the $H^+ h$ and $H^+ A$ production rates.
(We note that the tree level production rate of $H^+ A$ pair
in the MSSM is independent of $\tan \beta$.)

 As clearly
shown in Fig.~\ref{fig:cs}, the LHC can be sensitive to the entire
LHS parameter space, assuming that the above signal event
signature can be measured at the 1~fb level~\cite{Cao:2003tr}.
 The potential of the Tevatron to observe the
$H^+ A/H^+ h$ production processes deserves special investigation
and will be reported elsewhere~\cite{ahp-future}. We also note
that when $s_{\beta\alpha}$ is small, the tree level bottom and
$\tau$ Yukawa couplings are enhanced by a factor of
%cpy%
$(-\sin\alpha/\cos\beta)\simeq \tan\beta$, compared with the SM
values. Therefore, the LHS, which is realized in
intermediate-to-high $\tan\beta$ region, can be potentially probed
even at the Tevatron via several {\it $\tan\beta$-enhanced}
processes, such as $p\bar{p}\to h(A)$
(produced via gluon-gluon fusion process)
with $h/A\to\tau\bar{\tau}$, $p\bar{p}\to b\bar{b}
h(A)$,  as well as $p\bar{p}\to t\bar{t}$ with $t\to H^+ b$. At
present luminosity, those processes are sensitive only to very
large values of $\tan\beta\gtrsim 45-50$, while at $10$ fb$^{-1}$,
$\tan\beta\gtrsim 30$ could be
probed~\cite{Carena:2000yx,Belyaev:2002zz}. { At the LHC, a
smaller $\tan\beta$ value ($\gtrsim  10$) of the LHS can be tested
via the $\tan\beta$-enhanced processes, such as $pp\to (h,H,A) \to
\tau {\bar  \tau}$~\cite{Denegri:2001pn}. Furthermore, given the
expected large number ($\sim 10^8$) of top quark pairs produced at
the LHC, the LHS can manifest itself in the copious $t\to H^+ b$
decays as long as $M_{H^+}$ is not too large
(below about  $140$\,GeV)~\cite{Baarmand:2006dm}.
}%
\\
\\
%\input h+a_4conclusions.tex
%\section{Conclusions}
{\bf Conclusions:}
We have shown that in the MSSM the possibility
for the CP-even Higgs boson $h$ to be lighter than $Z$-boson
(as low as about 60 GeV) is, contrary to the usual belief,  not
yet excluded by the existing direct search experiments. The
characteristic of the light Higgs boson scenario (LHS) is that the
$ZZh$ coupling and the decay branching ratio ${\rm Br}(h/A\to
b\bar{b})$ are simultaneously suppressed as a result of SUSY loop
corrections. We would also  note that the region of MSSM
parameter space considered for explaining the non-conclusive LEP2
excess of the $\sim 98$~GeV 'Higgs-like'
events~\cite{Barate:2003sz}, as studied in the literature (see,
e.g.,~\cite{Drees:2005jg,Kim:2006mb}), is a subset of the more
generic LHS parameter space that we have found in this paper. Our
result would be useful for clarifying the parameter space
responsible for this excess.

The implications of the LHS to the usual LHC
(and Tevatron) search strategies for
the lighter CP-even Higgs boson ($h$) can be summarized as
follows. In view of its production mechanisms, both the vector
boson fusion process and the associated production of $h$ with
vector boson are largely suppressed, while the associated
production of $h$ and $H^+$ is enhanced by the large $W$-$h$-$H^+$
coupling. In view of its decay channels, the decay branching ratio
of $h$ into $b {\bar b}$ mode is reduced and the $\tau^+ \tau^-$
mode is enhanced. Also, as compared to the SM rates, the $gg \to h
\to \gamma \gamma$ rate is reduced by a couple of orders of
magnitudes and the $gg \to h \to \tau^+ \tau^-$ rate is enhanced
by about an order of magnitude for $h$ around 60 GeV. Since
the mass of the heavier CP-even Higgs boson is below 130
GeV in the LHS, there is no resonance enhanced $hh$ pair production from $gg$
fusion process.
The only large Higgs pair production rate at the
LHC is via $pp \to H^\pm h (A)$ whose production cross sections are
sizable (above a few hundreds fb) and insensitive to the value of 
$\tan \beta$. (The tree level $AH^\pm$ rate is 
independent of $\tan \beta$.)
Hence, if
this production channel is not observed at the LHC, it would undoubtedly
exclude the LHS. On the other hand,
if this production channel is detected, a
large production rate of the heavier Higgs boson $H$ via vector boson
fusion process is expected in the LHS.

Finally, we note that in the LHS, $B$ physics processes at
$B$-factories, Tevatron and LHC, such as $b\rightarrow s \gamma$, 
$B^-\rightarrow \tau^-
\bar{\nu}$, $B_{d,s}\rightarrow
\mu^+\mu^-$
and $B_s-\bar{B}_s$ oscillation
measurements, could be largely
modified due to the sizable contributions from light
(neutral and charged) Higgs bosons.
Since the predictions on those processes could strongly depend on
the flavor structure of the SUSY breaking parameters, we do not
impose any constraints from flavor physics to further restrict the
allowed MSSM parameter space of the LHS presented in this work. A
detailed study of the constraints from flavor physics, under a
specific assumption of the flavor structure, is interesting and
deserves a separate study.

\begin{figure}[htb]
\vspace*{-0.3cm}
\includegraphics[width=0.5\textwidth]{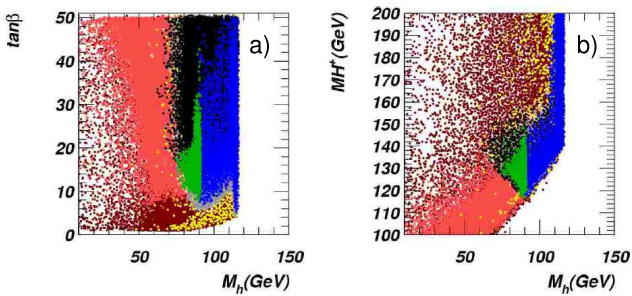}
\vspace*{-0.4cm}
\caption{\label{fig:bphys} The allowed LHS parameter space (green area)
after application  constraints from B-physics (black and gray areas)
combined with LEP2constraints (red, dark red and yellow areas).}
\vspace*{-0.2cm}
\end{figure}

Our preliminary study~\cite{ahp-future} shows that even for the commonly
discussed minimal-flavor-violation (MFV) scenario in which
flavor violation is solely generated by
SM Cabibbo-Kobayashi-Maskawa (CKM) matrix, the LHS can
be consistent with all the present $B$-physics data though
its parameter space is largely reduced.
In Fig.~\ref{fig:bphys} we present 
the allowed LHS parameter space indicated by green area
after application  B-physics constraints (black and gray areas)
combined with LEP2 constraints (red, dark red and yellow areas).
The blue color indicates the allowed parameter space for $m_h>M_Z$.
All other colors indicate the excluded regions.
One can see that $m_h$ is required to be larger than
about 80 GeV while  $\tan\beta$ is bounded to be less than about 20,
mainly due to the $B_{d,s}\rightarrow \mu^+\mu^-$ constraint
represented  by black area.
Hence, it is expected that the MSSM LHS would require a
non-MFV flavor sector, should $h$ boson be much lighter than $Z$ boson.
Its detail will be published in a forthcoming paper~\cite{ahp-future}.
\\
\\
{\bf Acknowledgments:}

A.~B. thanks SUSY 2007 organizers  for warm hospitality.
% BibTeX users please use
 \bibliographystyle{apsrev}
 \bibliography{h+a}

\end{document}